\documentclass[prb,twocolumn,showpacs,preprintnumbers,amsmath,amssymb]{revtex4}
\usepackage{graphicx}
\usepackage{dcolumn}
\usepackage{bm}

\begin{document}

\title{Electronic Structure Calculations with the Tran-Blaha
Modified Becke-Johnson Density Functional}

\author{David J. Singh}

\affiliation{Materials Science and Technology Division,
Oak Ridge National Laboratory, Oak Ridge, Tennessee 37831-6114}

\date{\today}

\begin{abstract}
We report a series of calculations testing the predictions of the
Tran-Blaha functional for the electronic structure and magnetic
properties of condensed systems.
We find a general improvement in the properties of semiconducting
and insulating systems, relative to calculations
with standard generalized gradient
approximations, although this is not always by the same mechanism
as other approaches such as the quasiparticle GW method.
In ZnO the valence bands are narrowed and the band gap is increased to
a value in much better agreement with experiment. The Zn $d$ states
do not move to higher binding energy as they do in LDA+U calculations.
The functional is effective for systems with hydride anions,
where correcting self-interaction errors in the 1$s$ state is important.
Similarly,
it correctly opens semiconducting gaps in the alkaline earth hexaborides.
It correctly
stabilizes an antiferromagnetic insulating ground state for the
undoped cuprate parent CaCuO$_2$, but seriously degrades the agreement with
experiment for ferromagnetic Gd relative to the standard local
spin density approximation and generalized
gradient approximations. This is
due to positioning of the minority spin $4f$ states
at too low an energy.
Conversly, the position of the La $4f$ conduction bands of La$_2$O$_3$
is in reasonable accord with experiment as it is with standard
functionals.
The functional
narrows the Fe $d$ bands of the parent compound LaFeAsO
of the iron high temperature superconductors, while maintaining the
high Fe spectral weight near the Fermi energy.
\end{abstract}

\pacs{}

\maketitle

\section{introduction}

Density functional theory (DFT) is formally a theory for the ground state
and not for the spectroscopic properties, at least not with the
interpretation of Kohn-Sham eigenvalues as a one electron spectrum.
In fact, DFT based methods, using modern generalized gradient approximations
\cite{gga,pbe}
(GGA)s give extremely useful total energies and structural properties
for a wide range of materials,
but is known to strongly underestimate band gaps in most semiconductors
and insulators. It also suffers from incorrect predictions of metallic
or very small band gap semiconducting states in Mott insulators, and in
the case of undoped cuprates (e.g.
La$_2$CuO$_4$ and CaCuO$_2$) produces an incorrect non-magnetic
ground state in contrast to the actual antiferromagnetic insulating
state.

An important aspect of standard DFT calculations is that they are
computationally efficient.
This allows detailed calculations for wide variety of
complex materials including
their full realistic chemical and
structural details.
This greatly facilitates comparison with experiment.
Similarly and in spite of the known problems with standard DFT calculations,
such calculations
have proved to be very useful in understanding electronic spectra
of solids, both through interpretation of experiments and predictions.
Methods that retain the computational efficiency of standard DFT,
but improve the spectroscopic properties are potentially of great
interest.
In this regard, Tran and Blaha have recently developed a
semi-local functional \cite{mbj} based on a modification of the
Becke-Johnson functional. \cite{bj}
Importantly, this is not a hybrid functional, and therefore
scales like standard DFT rather than Hartree-Fock in solids.
They showed that this functional gives very much improved band gaps
for a variety of insulators, including semiconductors, oxides,
rare gas solids and lithium halides. \cite{mbj} 
They also found substantial gaps in accord with experiment for the
3$d$ transition metal monoxides, MnO, FeO and especially the
prototypical Mott insulator NiO.
Calculations of the optical properties for a variety of heavier
halides also showed much improved agreement with available
experimental data. \cite{singh-halides}

Here we present further tests of this functional emphasizing systems
that are difficult to describe for a variety of different reasons.
We also discuss in more detail the electronic structure of ZnO, which
was one of the systems studied by Tran and Blaha.
We find general improvements in the band gaps over the standard PBE
functional in all materials. We also find an antiferromagnetic
insulating ground state for the Mott insulator CaCuO$_2$. However,
the electronic structure differs from that
predicted by other methods and its consistency with experimental
data as regards the distribution of the $d$ spectral weight is not 
clear. The
electronic structure of ferromagnetic Gd is degraded
with respect to the standard PBE functional.

\section{approach}

The calculations were performed with the standard general potential
linearized augmented planewave (LAPW) method with local orbitals,
\cite{singh-book,singh-lo} as implemented in the
WIEN2k package. \cite{wien}
We use the Perdew Burke Ernzerhof (PBE) GGA as a reference, with
the exception of CaCuO$_2$, where we use PBE+U calculations, with
Coulomb parameter $U$-$J$=7 eV and self-interaction correction
\cite{u-sic}
(SIC) and about mean field \cite{u-amf}
(AMF) double counting corrections, since standard GGAs do not
produce an antiferromagnetic ground state in that case. \cite{singh-cacuo2}
Except as noted, we used experimental lattice parameters and relaxed all
non-symmetry fixed internal coordinates with the PBE GGA.
Also, except as noted, we performed calculations in a scalar relativistic
approximation for the valence states, while core states were treated
relativistically.
The basis sets and zone samplings were well converged, with additional
local orbitals used to relax linearization errors.
We used the standard LAPW matching as opposed to the faster
augmented planewave plus local orbital (APW+lo) approach.
\cite{sjo}

\section{semiconductors and non-magnetic insulators}

\begin{figure}
\includegraphics[width=\columnwidth]{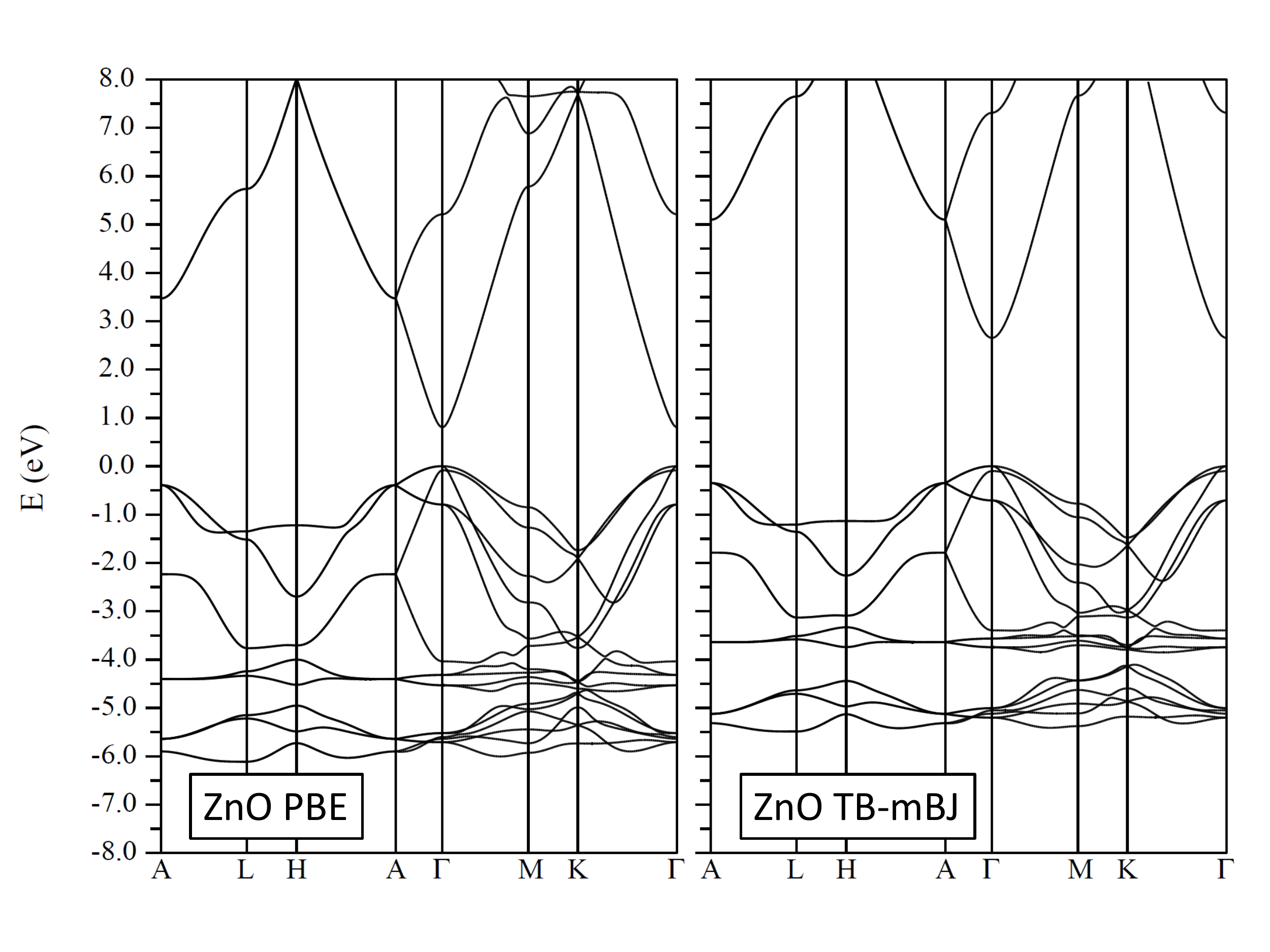}
\caption{
Comparison of the PBE and TB-mBJ band structures
for ZnO.}
\label{ZnO-bands}
\end{figure}

We begin with ZnO,
which is a well studied material with a band gap of 3.44 eV.
Standard functionals give an unusually large band gap error in this
material.
Correcting this error is of considerable importance in understanding
the defect physics and semiconducting properties of ZnO.
\cite{lany}
Our PBE calculations, which we performed with the experimental
lattice parameters of $a$=3.2489 \AA, $c$=5.2049 \AA, and internal
parameter $u$=0.381 yield a band gap of 0.80 eV. The band structure
is shown in Fig. \ref{ZnO-bands}.
The unusually large underestimate is generally thought to be
associated with a position of the Zn 3$d$ bands
at too low binding energy with respect to the valence band edge.
\cite{laskowski,shishkin}
In this case, the
resulting overly large
level repulsion between the $3d$ states and valence bands
pushes the valence band maximum up, reducing the band gap.
This view is supported by photoemission measurements, which place
the $3d$ states $\sim$ 7 -- 8 eV below the valence band edge,
\cite{powell,ozawa}
although it should be noted that this could possibly
reflect properties of the surface and not the bulk. Considering the
importance of this material it would be desirable to further check
the position of the Zn $d$ bands, for example by higher energy
photoemission or other spectroscopies.

\begin{figure}
\includegraphics[width=0.9\columnwidth]{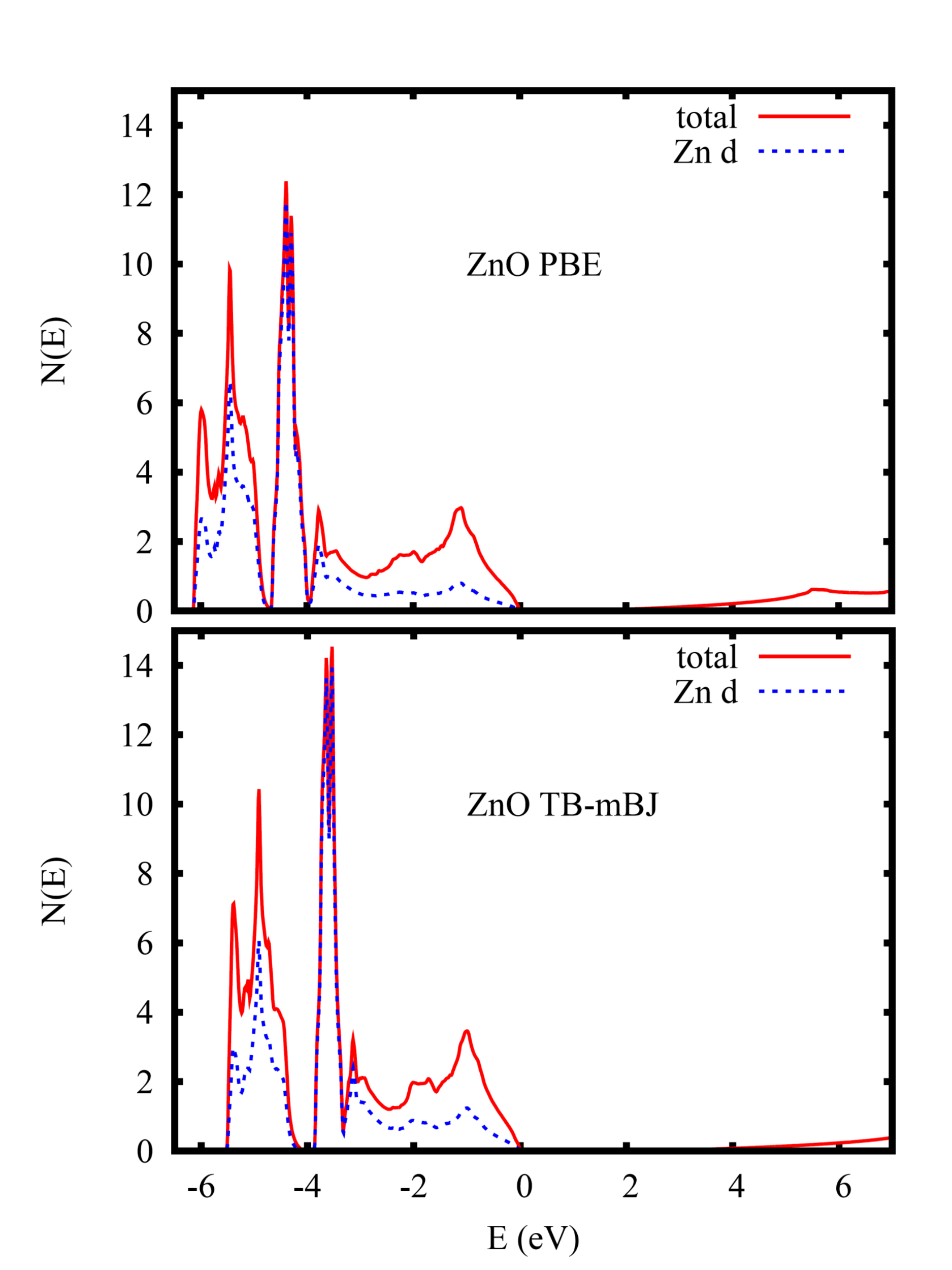}
\caption{(color online)
Comparison of the PBE (top) and TB-mBJ (bottom) electronic densities of states
for ZnO. The Zn $d$ character is by projection onto the Zn
LAPW sphere, radius 2.05 Bohr.}
\label{ZnO-dos}
\end{figure}

The band gap error is largely corrected by GW calculations.
\cite{vsg,shishkin}
These calculations place the Zn $d$ at high binding energy $\sim$ 6.5 eV,
as compared to $\sim$ -5 eV in standard PBE calculations (see
Fig. \ref{ZnO-dos}).
The error in the band gap is also greatly improved in LDA+U calculations,
where the Coulomb parameter is used to shift the Zn $d$ orbitals to higher
binding energy. \cite{laskowski}
Hybrid functional calculations also greatly improve the band
gap and shift the $d$ states to higher binding energy. \cite{uddin}
We obtained a band gap of 2.65 eV, which is much improved relative
to the PBE functional, but is still lower than experiment. This is almost
exactly the same value as that reported by Tran and Blaha.
\cite{mbj}
Our band structure (Figs. \ref{ZnO-bands} and
\ref{ZnO-dos}) shows Zn $d$ bands that are narrowed relative to the
PBE result, but are not shifted to higher binding energy.
This may well be the origin of the remaining discrepancy between the
TB-mBJ and experimental band gaps.

\begin{figure}
\includegraphics[width=\columnwidth]{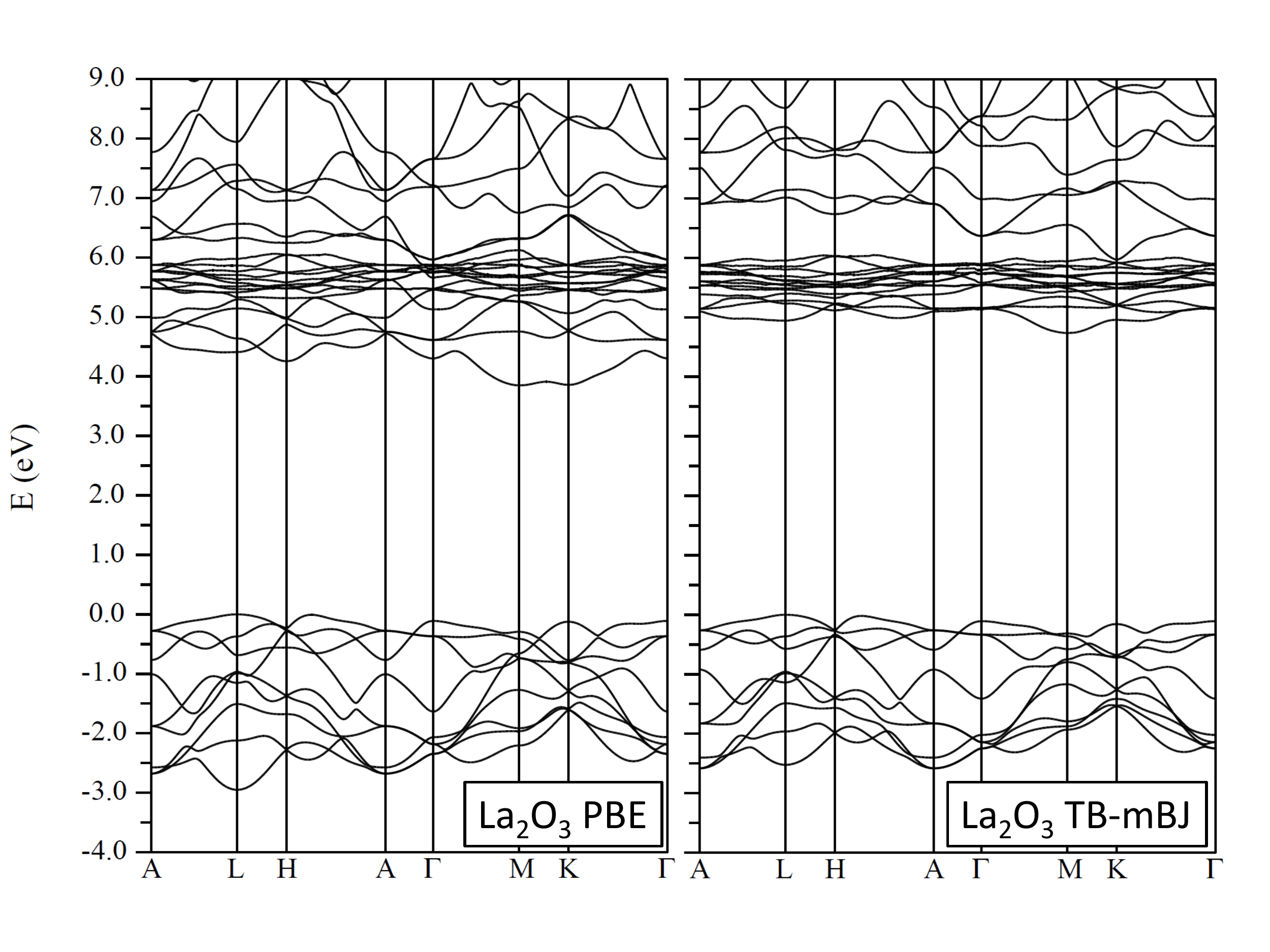}
\caption{
Comparison of the PBE and TB-mBJ band structures
for La$_2$O$_3$.}
\label{La2O3-bands}
\end{figure}

\begin{figure}
\includegraphics[width=0.9\columnwidth]{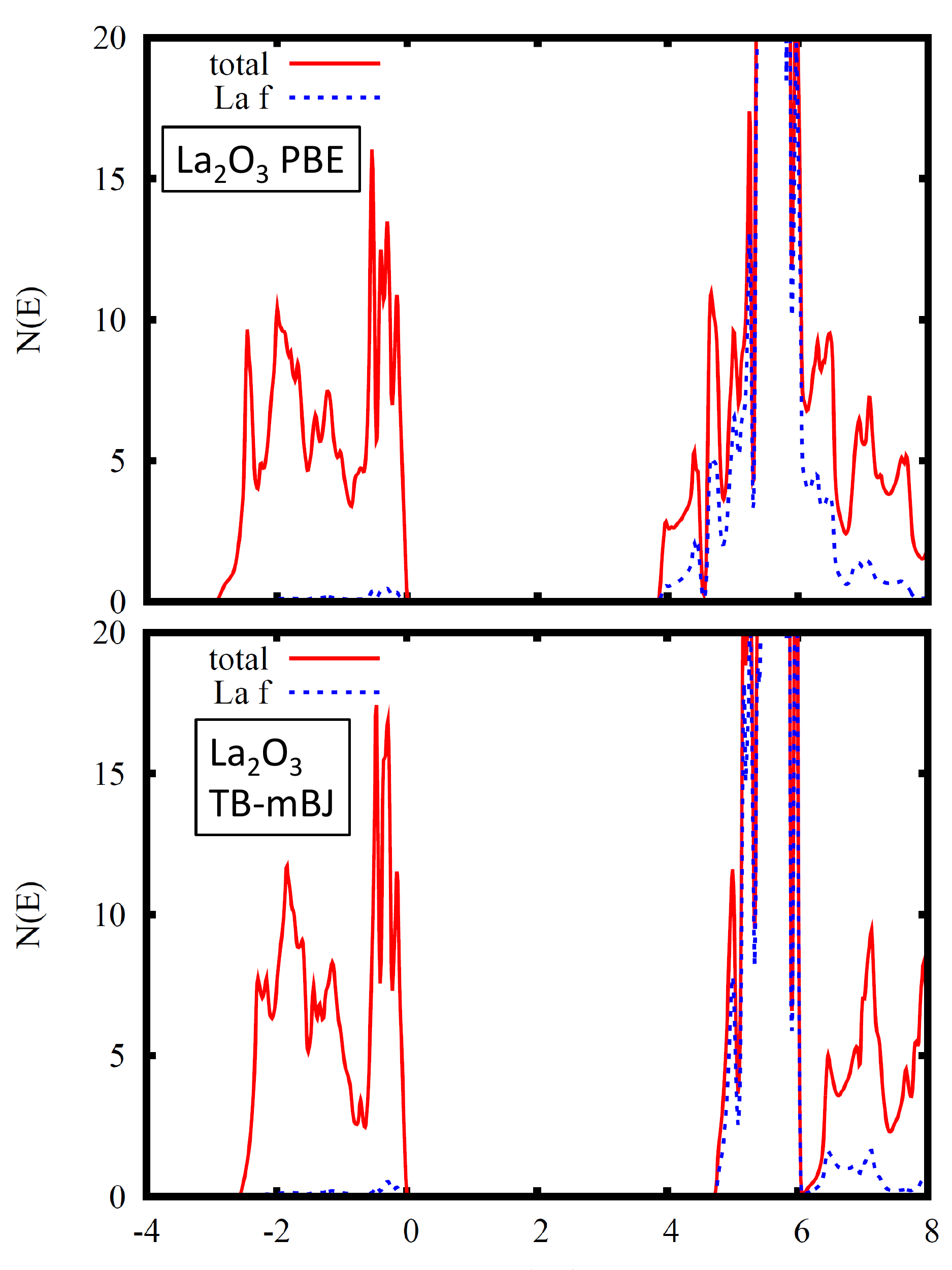}
\caption{(color online)
Comparison of the PBE (top) and TB-mBJ (bottom) electronic densities of states
for La$_2$O$_3$.  The La $f$ character is by projection onto the La
LAPW sphere, radius 2.5 Bohr.}
\label{La2O3-dos}
\end{figure}

We next discuss La$_2$O$_3$.
This high-k dielectric oxide occurs in a hexagonal ($P\bar{3}m1$ number
164) structure with lattice parameters $a$=3.943 \AA, and $c$=6.141 \AA.
Our calculated internal coordinates are $z_{\rm La}$=0.2468
and $z_{\rm O1}$=0.6458,
in good agreement with the experimental values of 0.2467 and 0.6470,
respectively.
\cite{bogatov}
La$_2$O$_3$ has a low temperature optical band gap of 5.34 eV.
\cite{kimura}
The calculated PBE and TB-mBJ band structures and densities of states (DOS)
are shown in Figs. \ref{La2O3-bands} and \ref{La2O3-dos}, respectively.
The valence bands are derived from O $2p$ states.
In the PBE calculation the conduction
bands come from La $sd$ states that hybridize with La $4f$ states that occur
just above the band edge, while in the TB-mBJ calculation the conduction bands
come from the La $4f$ states. Obtaining a correct band gap in this
compound depends on a correct placement of the La $4f$ resonance.
The calculated TB-mBJ band gap is 4.74 eV, which is higher
than the PBE value of 3.85 eV. It is in much better agreement with experiment
but still $\sim$ 0.6 eV too low.
Most likely this reflects the position of the La $4f$ resonance.
Interestingly, the position of the La $4f$ resonance with respect to the
O $2p$ derived valence band edge is almost the same in the PBE and TB-mBJ
calculations, although the $4f$ bands are narrower with the latter functional
because the other conduction bands are shifted to higher energy.
This behavior is reminiscent of what we found for La halides.
\cite{singh-halides}

\section{Ferromagnetic Gd}

\begin{figure}
\includegraphics[width=\columnwidth]{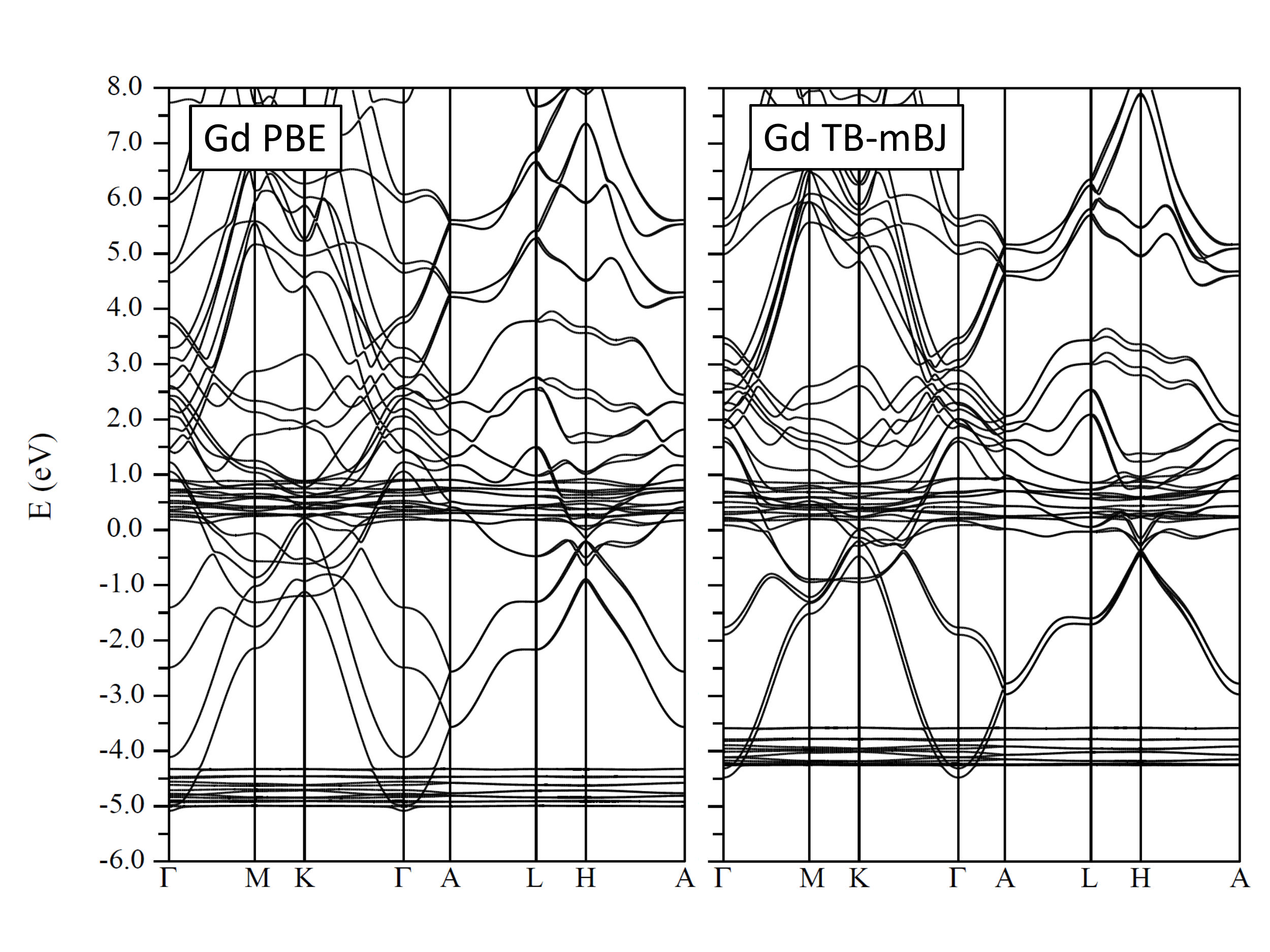}
\caption{Spin orbit split band structure of
ferromagnetic Gd with the PBE (left) and TB-mBJ (right) functionals.}
\label{Gd-bands}
\end{figure}

\begin{figure}
\includegraphics[width=\columnwidth]{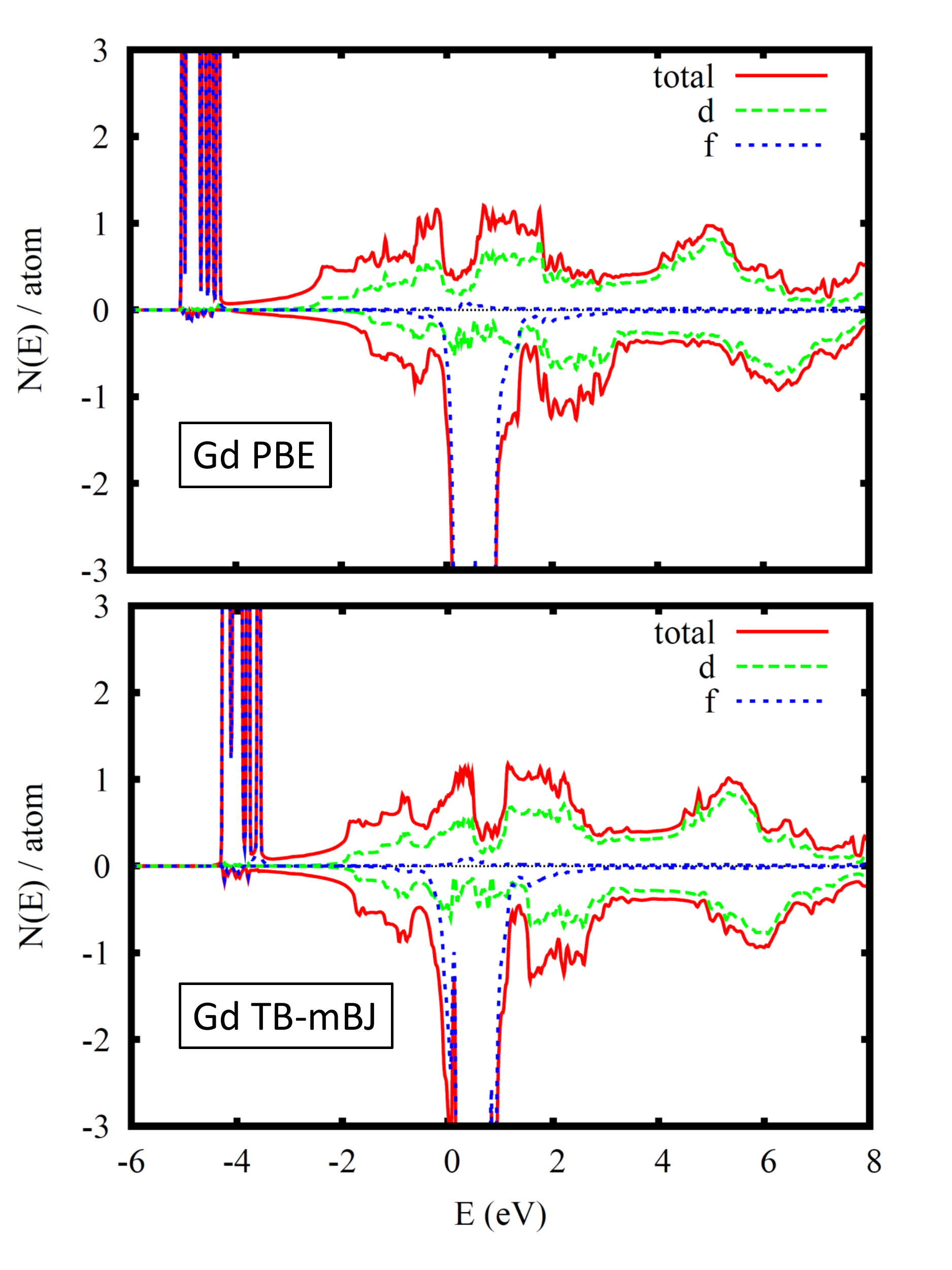}
\caption{(color online) Electronic DOS of ferromagnetic Gd.
Majority and minority spin projections are shown above and
below the horizontal axis, respectively. The orbital projections are onto the
Gd LAPW spheres of radius 3.1 Bohr.}
\label{Gd-dos}
\end{figure}

We next discuss an open $f$ shell material.
Gd is a ferromagnetic metal, with a half filled $4f$ shell. This leads
to a high spin moment, nominally 7 $\mu_B$ from the $4f$ electrons.
The actual experimental moment per atom
is 7.63 $\mu_B$, which includes a contribution
from polarization of the other valence states.
\cite{roeland,schutz,dowben}
The value of the moment is not sensitive to the treatment of the $f$
electrons provided that they remain fully polarized, as they do in standard
DFT and LDA+U calculations. \cite{singh-gd,abdelouahed}
On the other hand, standard DFT calculations produce a
Fermi energy density of
states, $N(E_F)$, that is higher
than that inferred from specific heat measurements. 
While there can be an enhancement of the specific heat in experiments
for a variety of reasons, such as electron-phonon interactions, spin
fluctuations, {\em etc.}, a reduction is difficult to understand.
The high $N(E_F)$ in calculations comes about because the
minority spin $f$ states, while above the Fermi energy, are close to it,
and hybridize with the valence bands.
LDA+U calculations move the minority bands to higher energy, reducing
the hybridization and improving agreement with experiment. In effect,
standard DFT calculations underestimate the exchange splitting of the
$4f$ shell.

We performed calculations for hcp
ferromagnetic Gd, including spin orbit
with the experimental lattice parameters, \cite{banister}
$a$=3.629 \AA, $c$=5.796 \AA.
We aligned the magnetization with the $c$-axis.

The band structures and DOS are shown in Figs. \ref{Gd-bands} and
\ref{Gd-dos}, respectively.
As may be seen, the TB-mBJ functional keeps the average $f$ position
approximately the same as in the PBE calculations, but reduces the
exchange splitting opposite to LDA+U calculations.
This brings additional $f$ character to the Fermi energy, reduces
the moment, and increases $N(E_F)$.
On a per atom basis we obtain $N(E_F)$=1.73 eV$^{-1}$ with the PBE
functional and 3.64 eV$^{-1}$ with the TB-mBJ functional.
The inferred experimental specific heat value is 1.57 eV$^{-1}$.
\cite{wells}
Our calculated PBE spin moment is 7.62 $\mu_B$ in agreement
with experiment and prior studies, while the TB-mBJ value
is 6.65 $\mu_B$.
Therefore, it may be concluded that the TB-mBJ functional seriously
degrades agreement with experiment for ferromagnetic Gd.

\section{hydrides}

Hydrides present an interesting challenge to DFT calculations, particularly
anionic hydrides.
Standard DFT calculations have large errors in their descriptions
of the one electron H atom because of an incomplete cancellation
of the Hartree and exchange correlation self-interactions.
The trihydride, YH$_3$, which is a transparent insulator,
is incorrectly described as a metal in the LDA.
This has been attributed to self-interaction on the H,
the ordinary problems in describing band gaps within the LDA,
other correlation effects,
and complications with the crystal structure.
\cite{alford,ng,ng1,geld,eder,geld2,miyake,wu,kelly,ahuja}
Several calculations have shown that a semiconducting gap can
be opened in GW quasiparticle calculations and also
the with the non-local weighted density approximation (WDA) to DFT.
\cite{wu}
The WDA is self-interaction free for one electron systems.
Unfortunately, while the insulating behavior of YH$_3$ is likely
an electronic effect and not a consequence of structural details, \cite{gogh}
the fact that the exact structure is uncertain complicates interpretation
of calculated results.

Here we present results for three well studied hydrides,
specifically LiH, MgH$_2$ and AlH$_3$.
LiH occurs in the fcc NaCl structure. Our calculations
were done with the experimental lattice parameter for
LiD at 83K, $a$=4.0447 \AA.
\cite{vidal}
MgH$_2$ is tetragonal, with a rutile structure.
We used the experimental lattice parameters, $a$=4.5198 \AA,
$c$=3.025 \AA. \cite{ono}
The internal H coordinate, $u$ was kept at its experimental value
of $u$=0.30478 since the force on H was found to be small with this
value.
AlH$_3$ occurs in a hexagonal, $R\bar{3}c$ structure. \cite{turley}
We used the experimental lattice parameters,
$a$=4.449 \AA, $c$=11.804 \AA, and relaxed internal coordinate. The
structure has Al on site 6$b$ (0,0,0) and H on site 18 $e$ ($x$,0,1/4),
with $x$=0.6319, in good accord with the reported experimental value of
$x$=0.628 (Ref. \onlinecite{turley}).

\begin{figure}
\includegraphics[width=\columnwidth]{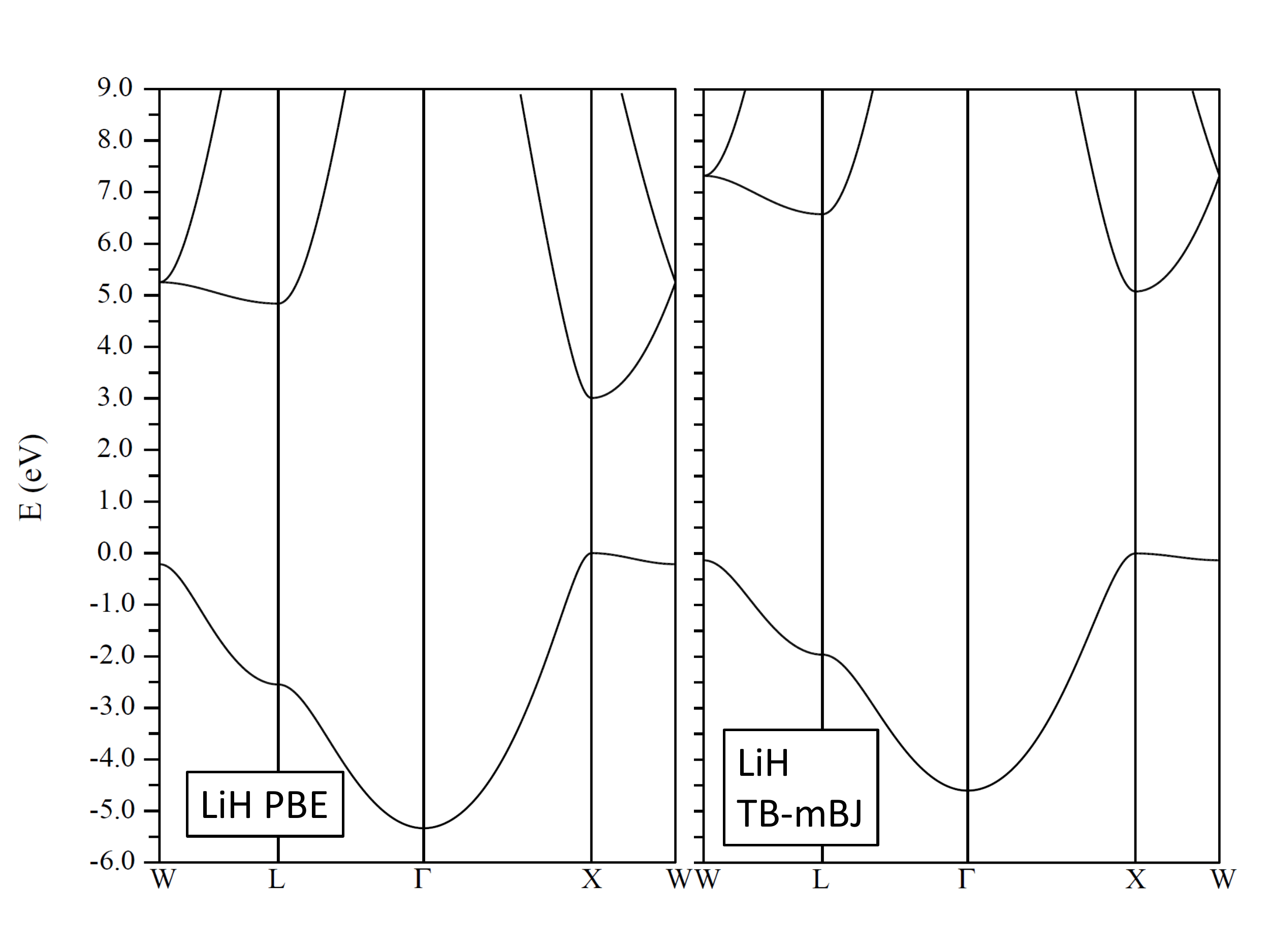}
\caption{
Band structure of LiH with the PBE (left) and TB-mBJ (right) functionals.}
\label{LiH-bands}
\end{figure}

\begin{figure}
\includegraphics[width=\columnwidth]{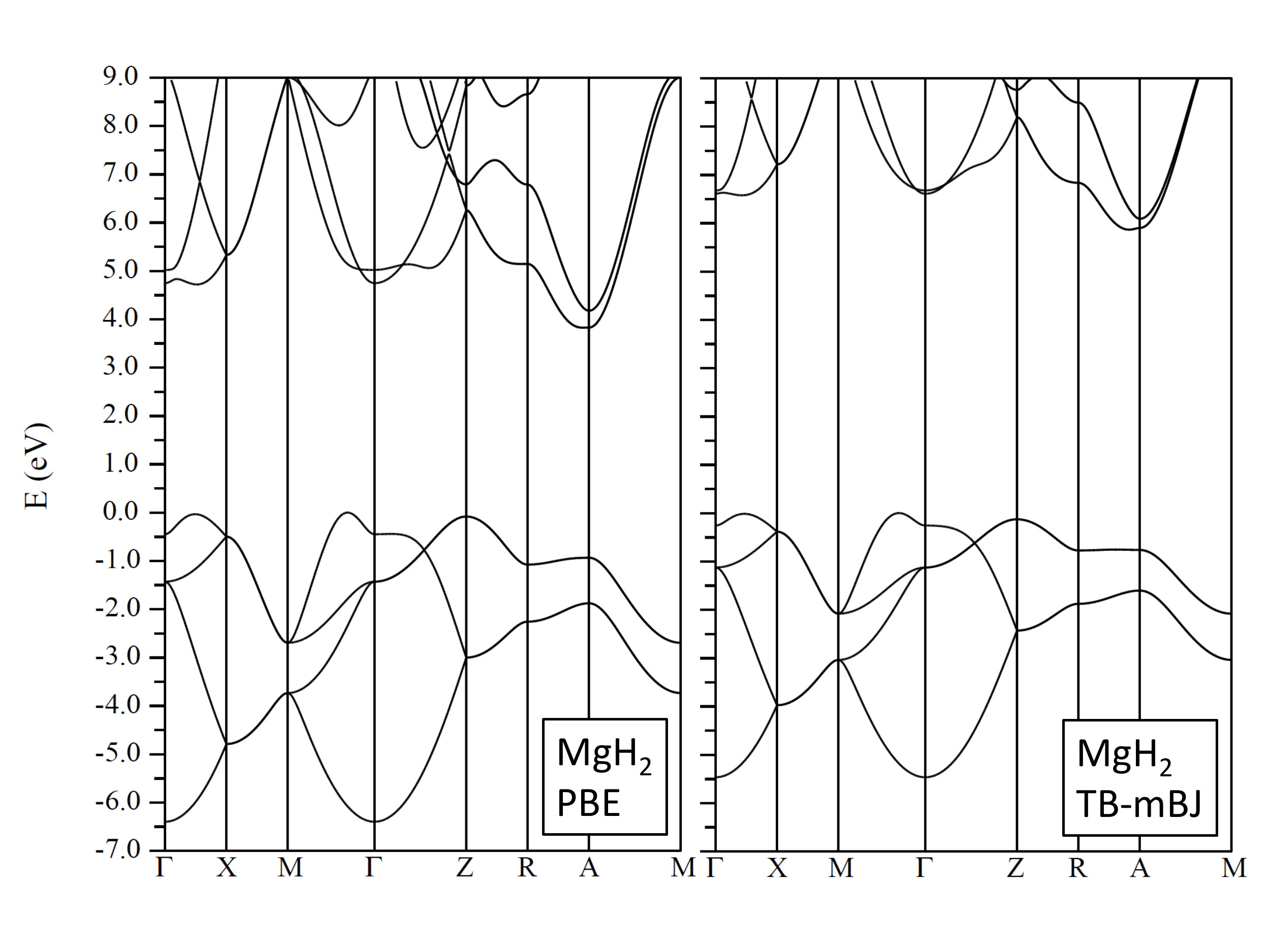}
\caption{
Band structure of MgH$_2$ with the PBE (left) and TB-mBJ (right) functionals.}
\label{MgH2-bands}
\end{figure}

\begin{figure}
\includegraphics[width=\columnwidth]{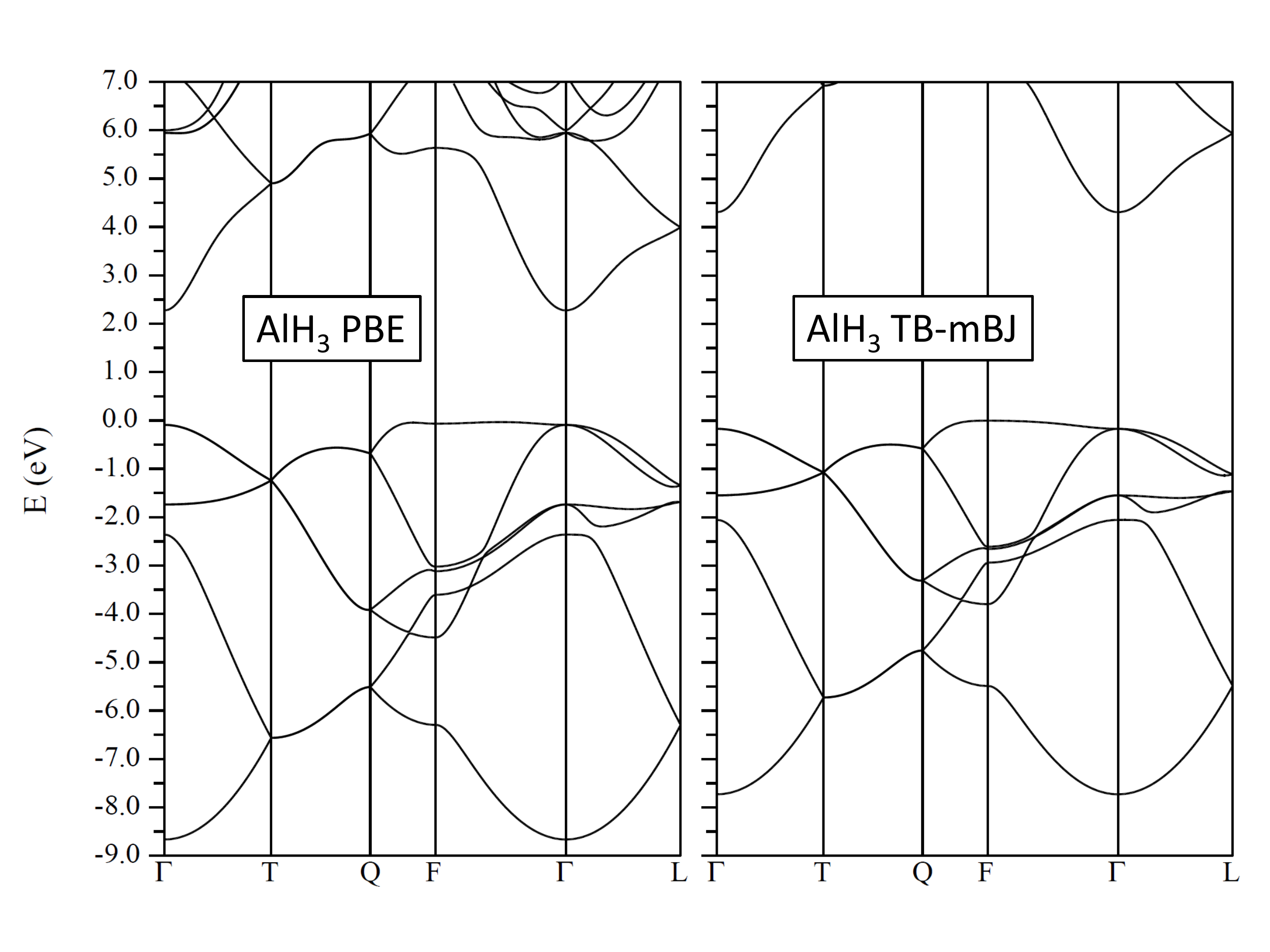}
\caption{
Band structure of AlH$_3$ with the PBE (left) and TB-mBJ (right) functionals.}
\label{AlH3-bands}
\end{figure}

\begin{figure}
\includegraphics[width=0.9\columnwidth,angle=0]{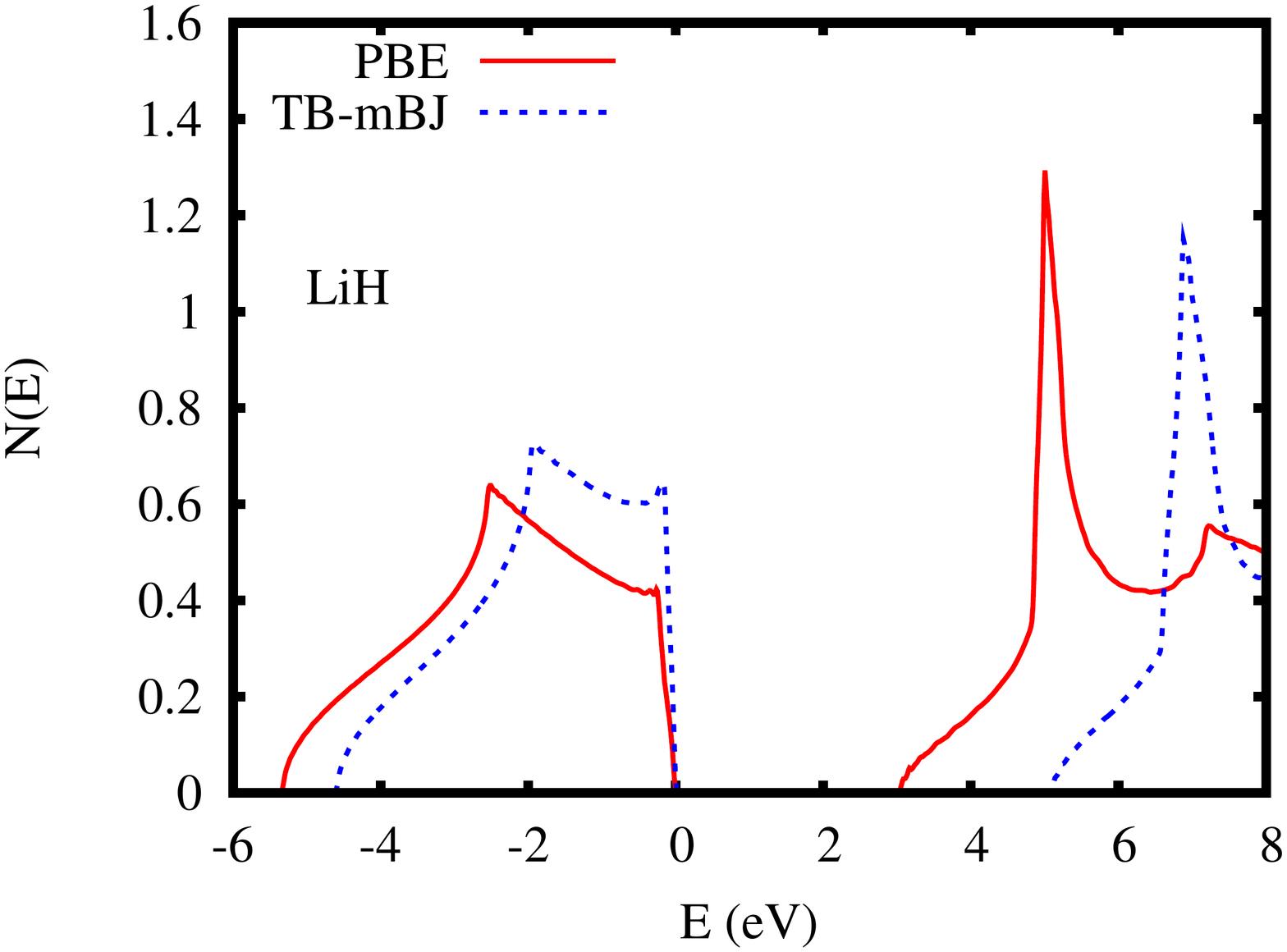}
\caption{(color online)
Calculated electronic DOS of LiH with the PBE and TB-mBJ functionals.}
\label{LiH-dos}
\end{figure}

\begin{figure}
\includegraphics[width=0.9\columnwidth,angle=0]{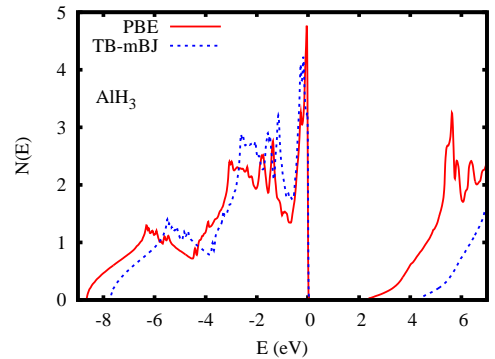}
\caption{(color online)
Calculated electronic DOS of AlH$_3$ with the PBE and TB-mBJ functionals.}
\label{AlH3-dos}
\end{figure}

Our calculated PBE and TB-mBJ band structures are given in Figs.
\ref{LiH-bands}, \ref{MgH2-bands} and \ref{AlH3-bands},
for LiH, MgH$_2$ and AlH$_3$, respectively.
The corresponding DOS are in Figs. \ref{LiH-dos} and \ref{AlH3-dos}
for LiH and AlH$_3$.
Our PBE band structures are similar to those obtained in prior DFT
calculations.
\cite{araujo,setten,lebegue,karazhanov}
The band gaps are of charge transfer character between predominantly
H $s$ derived valence bands and predominantly metal derived conduction bands.

We start with LiH. Both functionals predict a direct gap at the $X$ point.
The calculated band gaps are 3.01 eV and 5.08 eV for the PBE and TB-mBJ
functionals, respectively.
The experimental band gap of LiH is 5.0 eV as quoted in Ref. 
\onlinecite{setten}.
Thus the TB-mBJ is in almost perfect agreement with experiment for LiH.
MgH$_2$ has an indirect gap.
The PBE and TB-mBJ values are 3.70 eV and 5.74 eV, respectively.
The experimental band gap of MgH$_2$ is 5.6 eV.
\cite{isidorsson}
Therefore, as in LiH, almost perfect agreement with experiment is obtained
with the TB-mBJ functional.
We are not aware of an experimental band gap for AlH$_3$, but
recent GW calculations yield a values of 3.54 eV or 4.31
eV depending on the treatment. \cite{setten}
Our values are 2.27 eV and 4.31 eV for the PBE and TB-mBJ functionals.
In the cases studied the valence bands are somewhat narrower with the TB-mBJ
functional relative to the PBE results.
The results for hydrides with the TB-mBJ
functional are very encouraging.

\section{CaB$_6$ and SrB$_6$}

The hexaborides CaB$_6$ and SrB$_6$ are modest
band gap semiconductors.
\cite{souma,denlinger,cho,kim}
Like the anionic hydrides they have charge transfer gaps between
valence bands derived from a compact
orbital where self-interactions may be important and conduction bands
with metal atom character.
In this case the valence bands are derived from B $2p$ states.
Also, similar to YH$_3$, standard DFT calculation predict a semimetallic
state due to a band overlap, while both WDA and GW calculations yield
small insulating gaps.
\cite{tromp,kino,wu-cab6}

The nature of the bands is, however, different from the hydrides.
In the hydrides the valence bands are nominally from the
filled H $1s$ state, while the conduction bands are nominally from the
metal orbitals, although there may be some hybridization.
In the hexaborides, the bonding of the B network plays a central role.
The valence bands are derived from bonding combinations of B $p$ states,
while the conduction bands come from mixtures of non-bonding and antibonding
B $p$ states with Ca/Sr $d$ states. The center of the metal $d$ states
is at $\sim$4 eV above the band gap, but the hybridization of B $p$ and
Ca/Sr $d$ orbitals persists down to the conduction band minimum.
The band gap underestimation could therefore come from incomplete cancellation
of self-interaction errors in the B $p$ derived valence bands, or
alternatively an
overestimate of the B $p$ - Ca/Sr $d$ hybridization that then pushes the
conduction band minimum down (analogous to the pushing up of the valence
band edge in ZnO), or a combination of these two possible errors.

\begin{figure}
\includegraphics[width=\columnwidth]{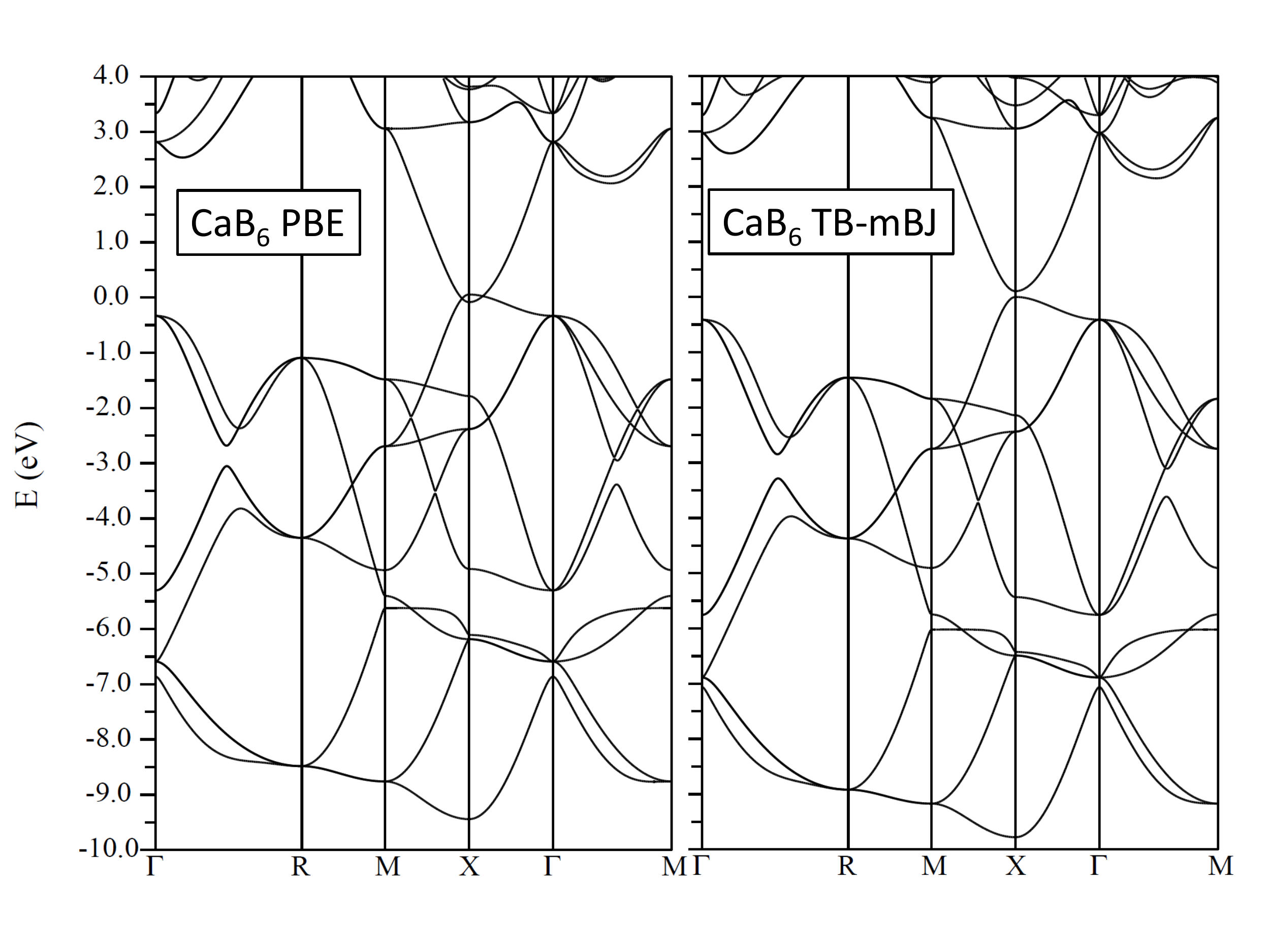}
\caption{
Band structure of CaB$_6$ with the PBE (left) and TB-mBJ (right) functionals.}
\label{CaB6-bands}
\end{figure}

\begin{figure}
\includegraphics[width=\columnwidth]{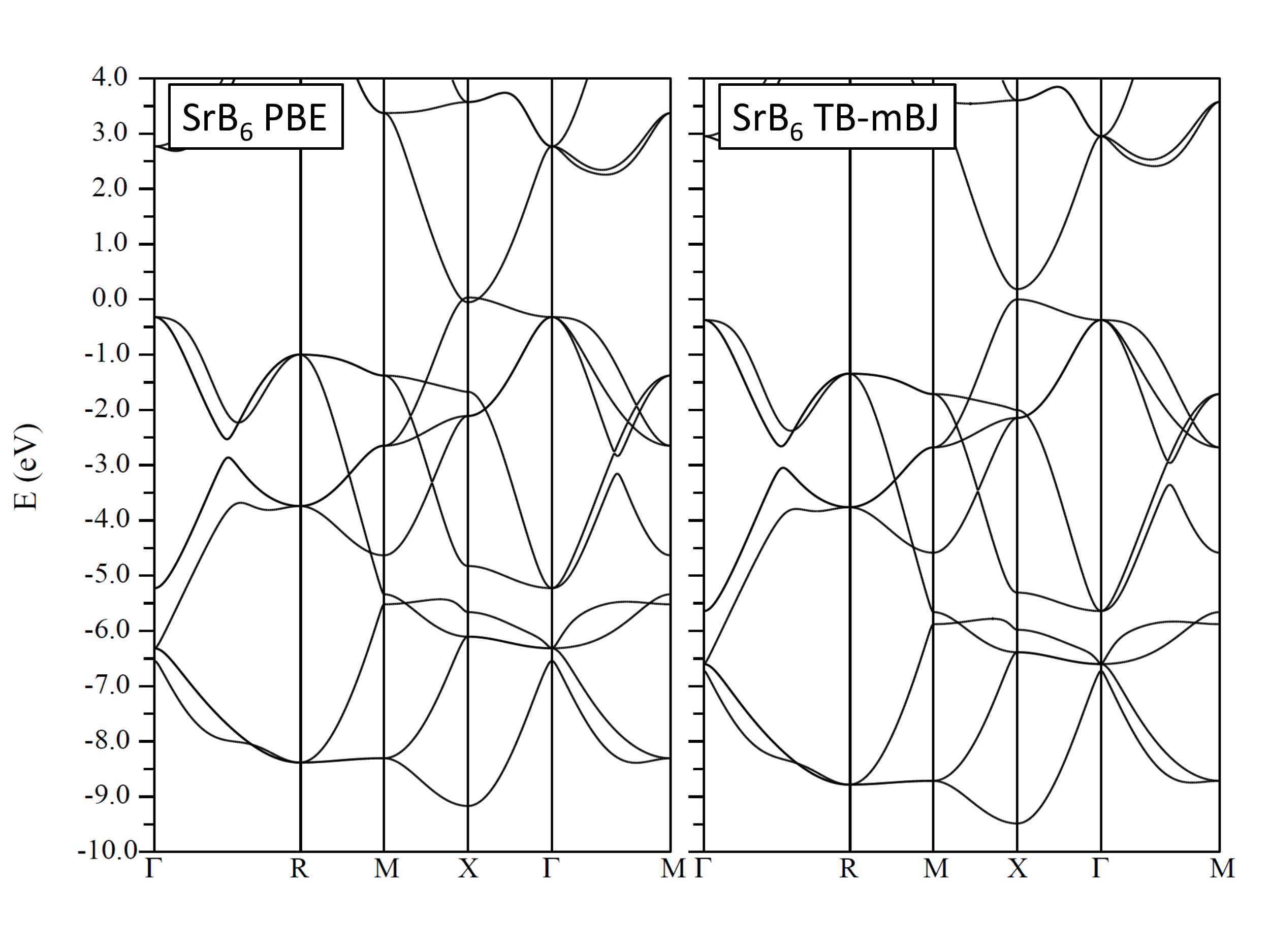}
\caption{
Band structure of SrB$_6$ with the PBE (left) and TB-mBJ (right) functionals.}
\label{SrB6-bands}
\end{figure}

\begin{figure}
\includegraphics[width=0.9\columnwidth,angle=0]{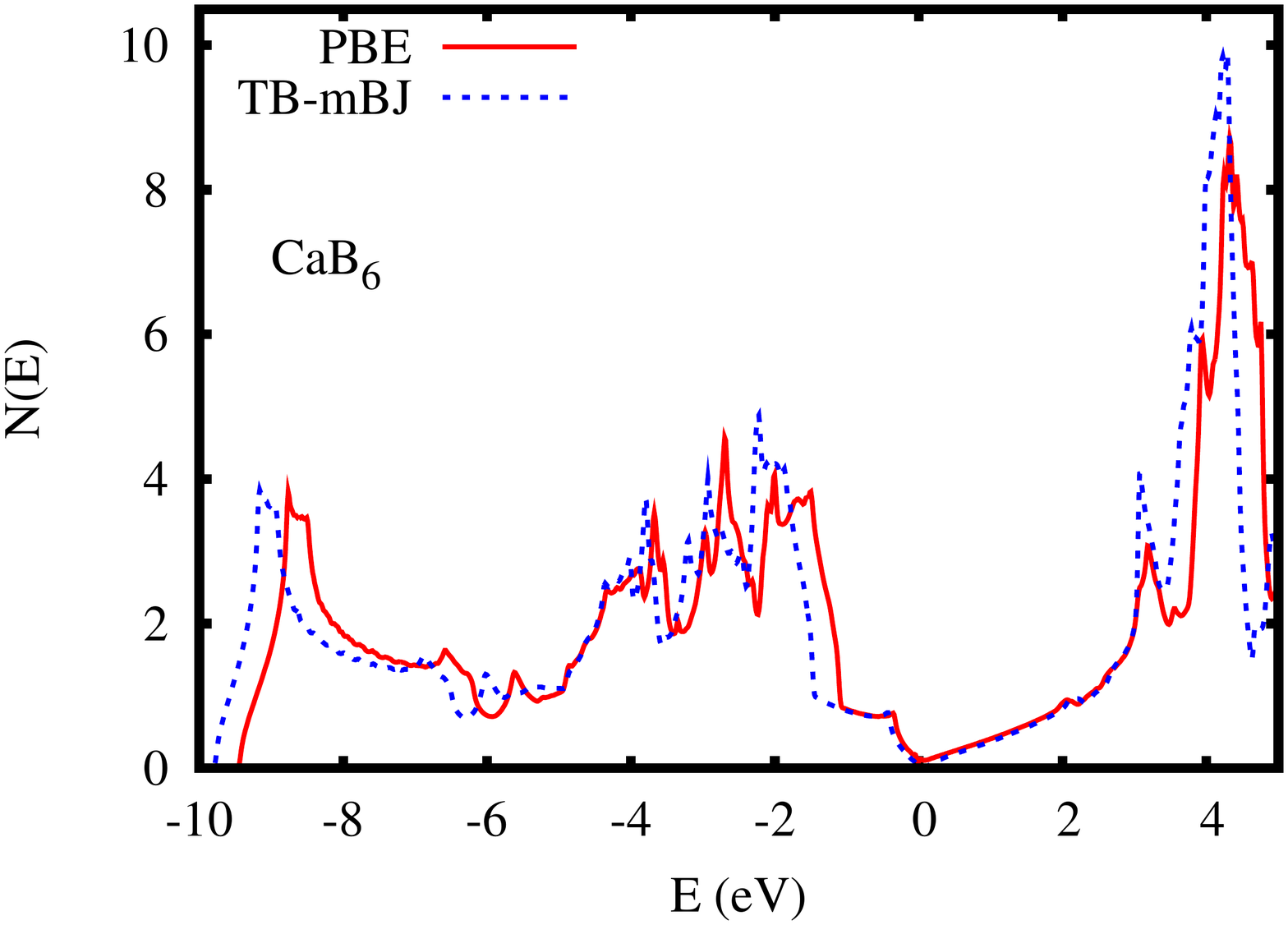}
\caption{(color online)
Calculated electronic DOS of CaB$_6$ with the PBE and TB-mBJ functionals.}
\label{CaB6-dos}
\end{figure}

\begin{figure}
\includegraphics[width=0.9\columnwidth,angle=0]{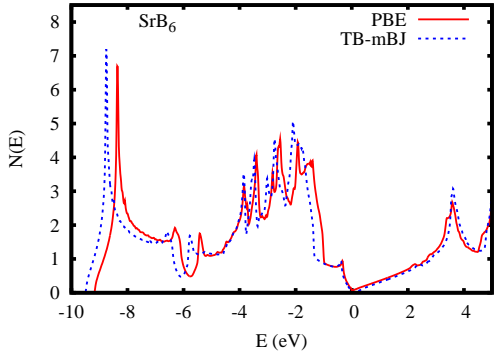}
\caption{(color online)
Calculated electronic DOS of SrB$_6$ with the PBE and TB-mBJ functionals.}
\label{SrB6-dos}
\end{figure}

We did calculations for CaB$_6$ and SrB$_6$ using the experimental simple
cubic crystal structures. We did not relax the internal B coordinate
because the forces with the experimental values were small.
The lattice parameters used were 4.1514 \AA, for CaB$_6$,
and 4.1953 \AA, for SrB$_6$. The internal coordinates were $x$=0.2019
for CaB$_6$ and $x$=0.2031 for SrB$_6$.
Our calculated band structures for CaB$_6$ and SrB$_6$ are given
in Figs. \ref{CaB6-bands} and \ref{SrB6-bands}, respectively.
The corresponding DOS are in Figs. \ref{CaB6-dos} and \ref{SrB6-dos}.

We find a metallic state in PBE calculations due to band
overlap at the $X$ point
in both materials.
On the other hand, with the TB-mBJ functional we obtain small gaps:
0.10 eV for CaB$_6$ and 0.18 eV for SrB$_6$.
There is a wide range in measured band gap values for the hexaborides.
However, recent optical measurements for CaB$_6$ yield a band gap of 0.25 eV,
\cite{kim}
while prior measurements indicate a larger gap $\sim$1 eV.
\cite{souma,denlinger,cho}
Considering the $T$ dependent resistivity, and the spectroscopic measurements,
the TB-mBJ result is clearly better than that of the PBE calculation, but
the gap is underestimated.
In contrast to the hydrides, we do not find a
valence band narrowing with the TB-mBJ functional.
Also, the position of the Ca/Sr $d$ states in the DOS (not shown) at $\sim$4
eV is almost exactly the same in the PBE and TB-mBJ calculations.
It will be
desirable to check this against spectroscopic data, if such data becomes
available.
An upward shift of the $d$ bands would increase the band gap.

\section{CaCuO$_2$}

As mentioned, Tran and Blaha reported calculations for the transition
metal monoxides, MnO, FeO and NiO finding improved band
gaps along with changes in the $d$ density of states of NiO that
differ from those found in LDA+U calculations. \cite{mbj}
Here we present calculations for
CaCuO$_2$. This material
is the simplest parent compound representative of undoped
cuprate superconductors.

These materials are Mott insulators when undoped, and become
high temperature superconductors when doped.
Importantly, they have a non-trivial band structure associated with
strong hybridization between Cu $d$ orbitals, particularly the $\sigma$
bonding Cu $d_{x^2-y^2}$ - O $p$ combinations. These combinations lead
to an antibonding band, of $d_{x^2-y^2}$ orbital symmetry crossing the Fermi 
energy and giving rise to the Fermi surface on which superconductivity
occurs. This hybridization also leads to a large interatomic exchange
coupling, and high magnetic energy scale, which may be important for
the high superconducting critical temperatures. When undoped, this band
is half filled, and as mentioned the Fermi surface is destroyed in favor of
a Mott insulator. While the Fermi surfaces of the doped phases are
generally well described in DFT calculations, the magnetism in these
materials is not. \cite{singh-cacuo2,pickett-htc}
In contrast to MnO, FeO and NiO, these materials
are described as non-magnetic in DFT calculations, and so the ground
state is qualitatively wrong, not only in terms of electronic structure
but also in the fact that there is no moment formation.
This can be corrected by LDA+U (or GGA+U) methods, which yield
an insulating antiferromagnetic ground state. \cite{anisimov}
These methods also shift the $d$ spectral weight away from the
Fermi energy to form what may be termed upper and lower Hubbard
bands. This shift of $d$ spectral weight away from the Fermi energy
is consistent with photoemission measurements.
\cite{damascelli}
However, while there is a shift towards Hubbard bands at high
binding energy in both doped and undoped cuprates, there does
remain $d$ spectral weight at the band edge in the insulators
and Fermi energy in the superconductors.

We performed calculations using the PBE+U method, with $U$-$J$=7 eV
and two double counting corrections, the self-interaction correction (SIC)
scheme and the about mean field (AMF) scheme.
We used the experimental tetragonal crystal structure (there
are no internal coordinates).
\cite{karpinski}
For the magnetic order we assumed a G-type ground state, i.e. the
in-plane nearest neighbor
checkerboard antiferromagnetism that is common to cuprates with
antiferromagnetic stacking along the $c$-axis.

\begin{figure}
\includegraphics[width=\columnwidth]{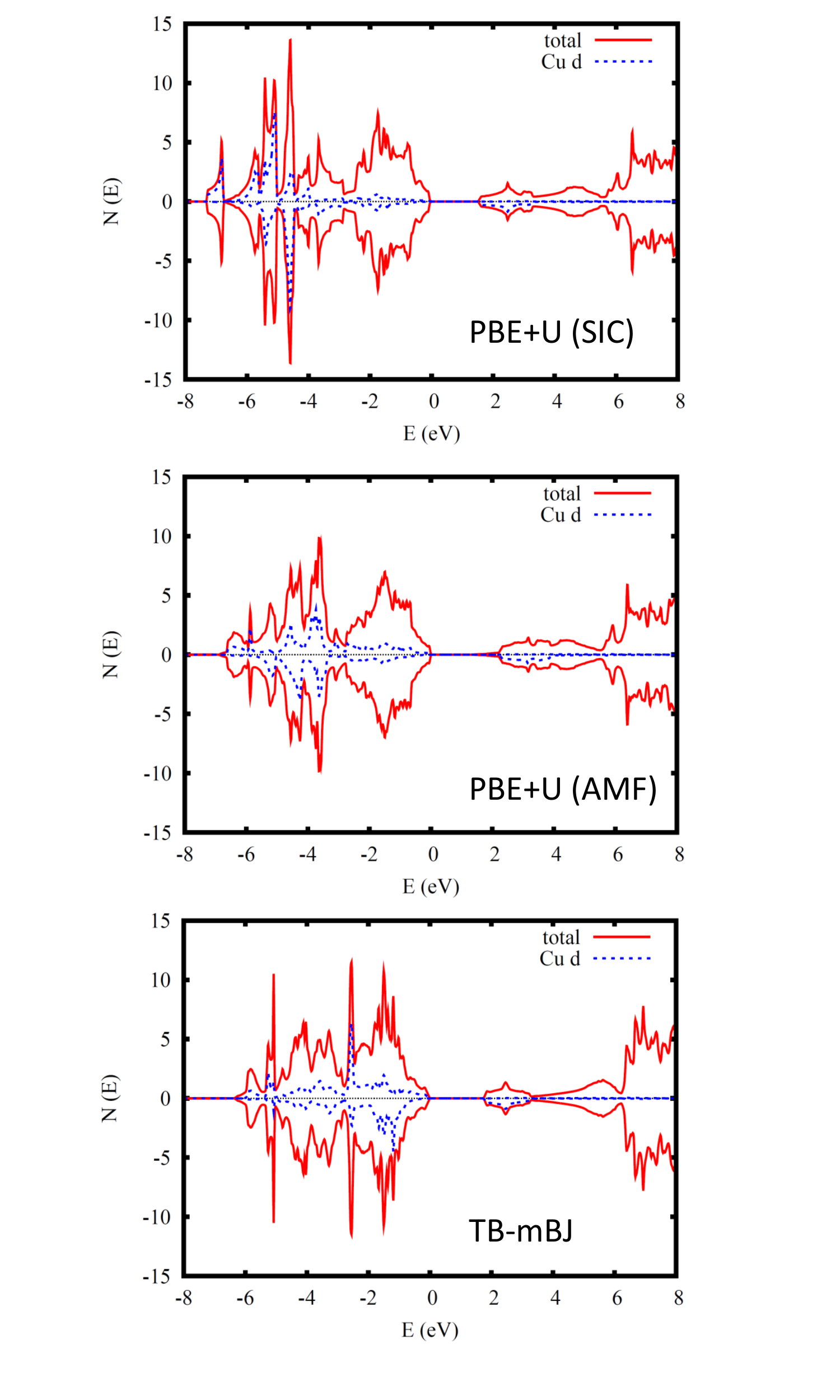}
\caption{(color online) Electronic DOS and projections for antiferromagnetic
CaCuO$_2$ with the PBE+U method (see text) and the TB-mBJ functional.
The Cu $d$ projection is onto the Cu LAPW sphere of radius 2.0 Bohr.
Majority and minority spin projections are shown above and below the
horizontal axis, respectively.
}
\label{CaCuO2-dos}
\end{figure}

The resulting electronic DOS are compared with results obtained
using the TB-mBJ
functional (and no added $U$) in Fig. \ref{CaCuO2-dos}.
First of all we note that unlike standard DFT treatments
the TB-mBJ functional successfully stabilizes
the antiferromagnetic insulating state.
The calculated spin moments as measured by the polarization inside the
Cu LAPW sphere (radius 2.0 Bohr) are 0.636 $\mu_B$, 0.695 $\mu_B$,
and 0.646 $\mu_B$ for the PBE+U (SIC), PBE+U (AMF) and TB-mBJ functionals,
respectively. The reduction from the ideal spin 1/2 value of 1 $\mu_B$ reflects
hybridization with the O $p$ states.
The calculated band gaps are sensitive to the value of $U$-$J$ assumed
in the PBE+U calculations. With the value of 7 eV used here, we obtained
gaps of
1.5 eV and 1.3 eV with the SIC and AMF double counting schemes, respectively.
The TB-mBJ gap is comparable at 1.7 eV.
The essential difference between the PBE+U and TB-mBJ results is seen
in the distribution of the Cu $d$ DOS. The two double counting schemes,
yield somewhat different distributions of the $d$ DOS, but both have
a strong shift of the occupied $d$ DOS to higher binding energy.
The TB-mBJ DOS instead places substantial $d$ DOS near the valence band
edge.
This is similar to what Tran and Blaha found in NiO.
In cuprates, it is clear that there are Hubbard bands at high binding energy,
which are present in the PBE+U but not the TB-mBJ
calculations. However, it is unclear from
existing experimental data whether the description near the valence
band edge is better with the PBE+U or the TB-mBJ functionals.
One possible avenue for resolving this is to note that when projected
onto a given site, the Cu $d$ character near the band edge with the
TB-mBJ functional has a substantial spin polarization, which could
in principle be detected in spin polarized scanning probe experiments.

\section{LaFeAsO}

\begin{figure}
\includegraphics[width=\columnwidth]{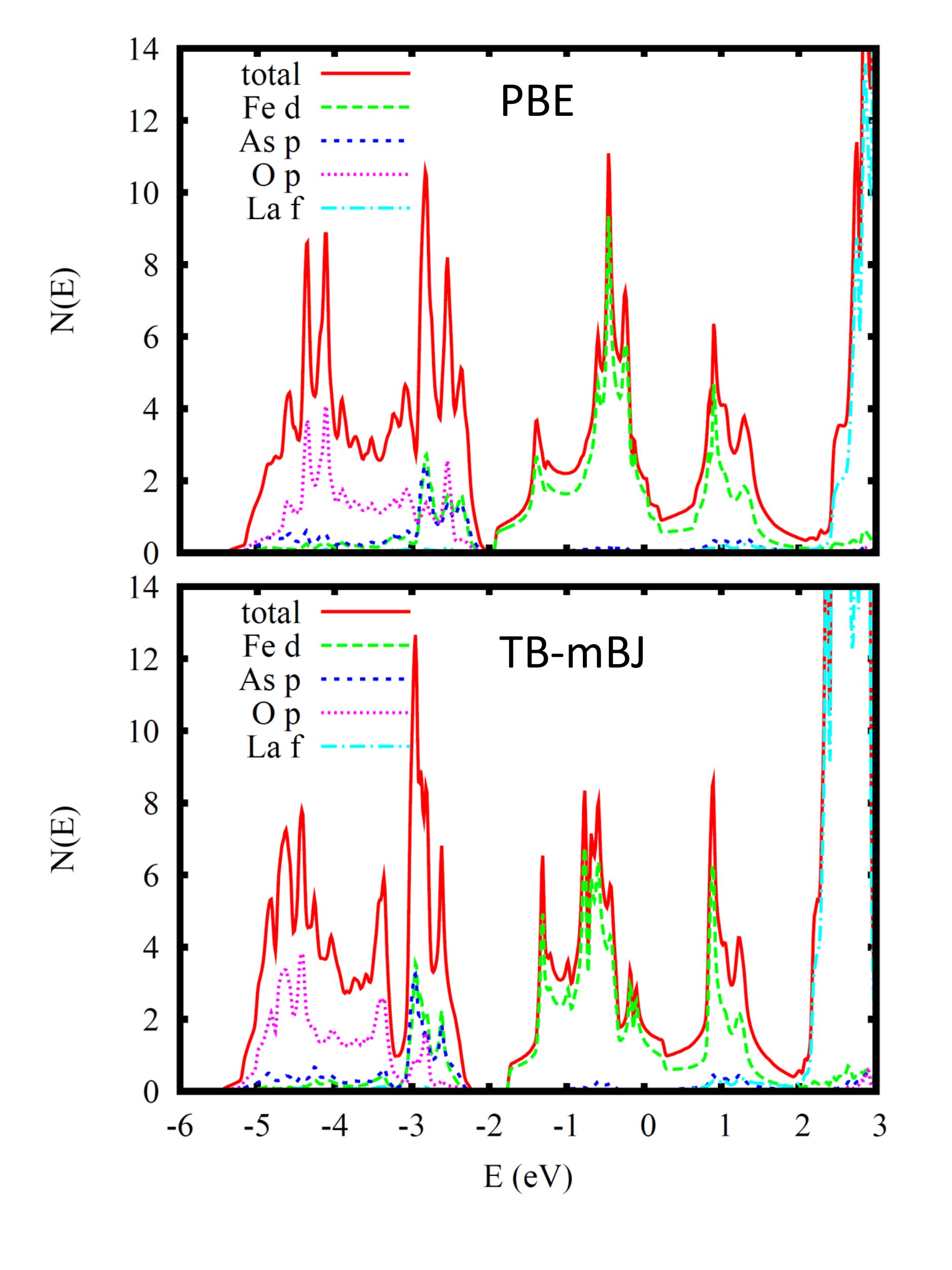}
\caption{(color online) Electronic DOS and projections for
LaFeAsO with the PBE and TB-mBJ functionals.
The projections are onto the LAPW spheres, of radii 2.5 Bohr,
2.25 Bohr, 2.25 Bohr and 1.75 Bohr, for La, Fe, As and O, respectively.}
\label{LaFeAsO-dos}
\end{figure}

\begin{figure}
\includegraphics[width=\columnwidth]{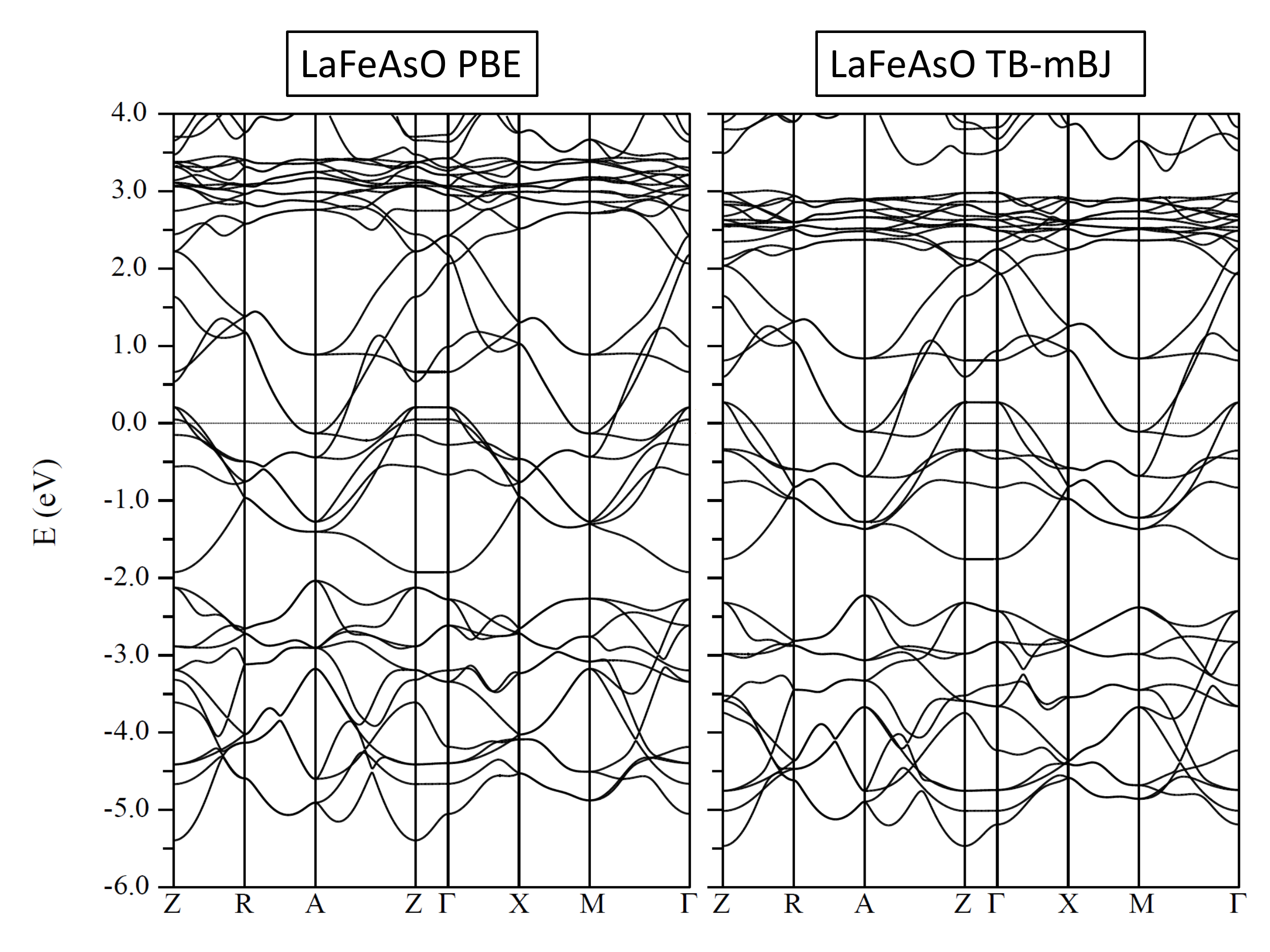}
\caption{
Band structure of LaFeAsO
with the PBE (left) and TB-mBJ (right) functionals.}
\label{LaFeAsO-bands}
\end{figure}

\begin{figure}
\includegraphics[width=\columnwidth]{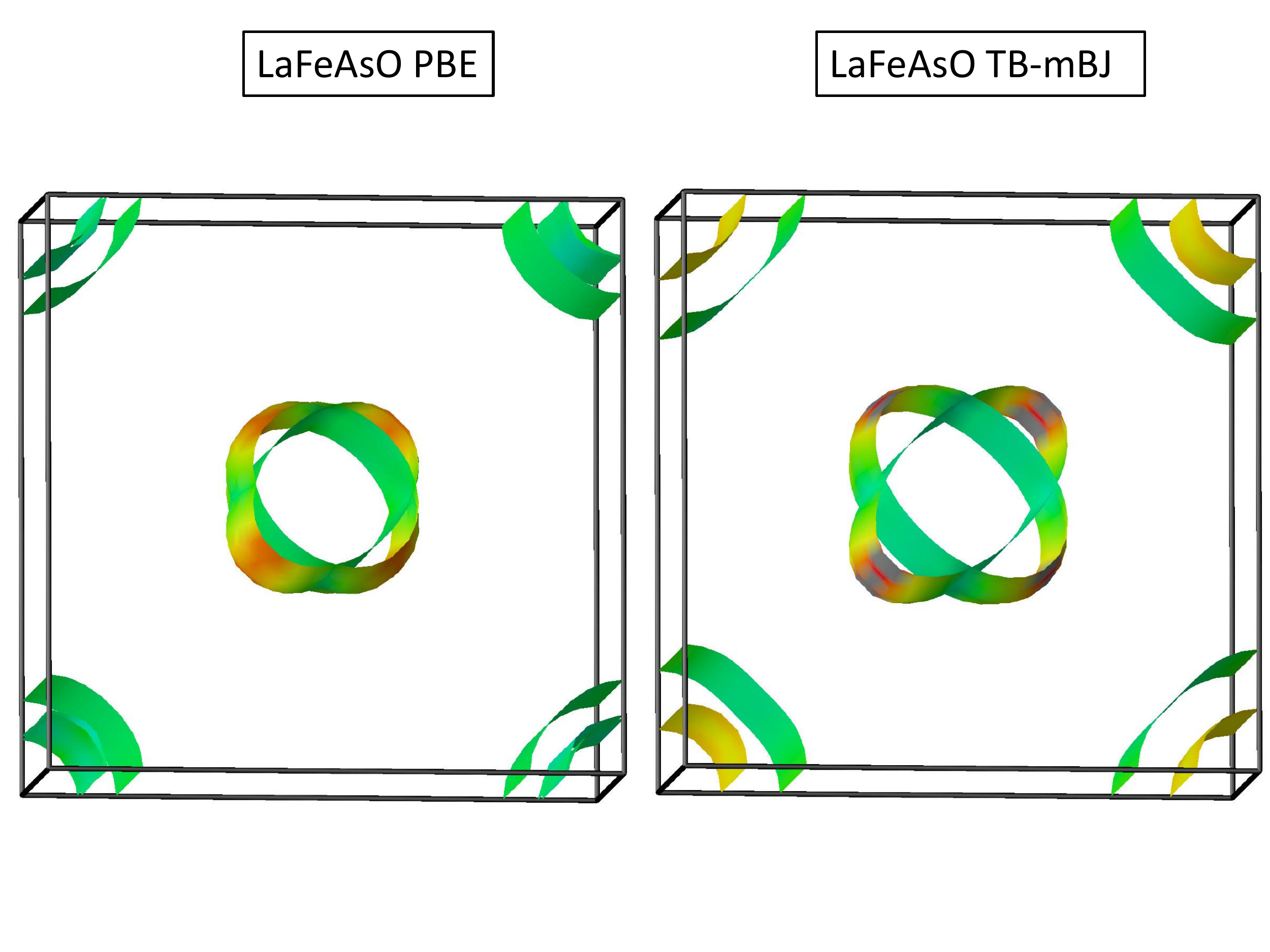}
\caption{(color online) Fermi surface of
LaFeAsO with the PBE (left) and TB-mBJ (right) functionals.
The shading is by band velocity and the corners are $\Gamma$
points.}
\label{LaFeAsO-fs}
\end{figure}

The iron based superconductors provide an interesting
contrast to the cuprates. \cite{kamihara}
Unlike cuprates, these materials show strong $d$ spectral weight
around the Fermi energy, in accord with DFT calculations, weak to
moderate correlations, and no Hubbard bands. \cite{lu-pe}
DFT calculations, \cite{singh-du}
show an Fe $d^6$ metal, with a high density of states and a
semimetallic character, i.e. small Fermi surfaces: hole-like sheets around
the zone center, and two electron sheets at the zone corner.
These features are in accord with experiment, and are degraded
by methods that build in strong correlations.
One difference between DFT calculations and experiment is that the
$d$ bands are narrower in experiment.
\cite{lu-pe,qazilbash}

We performed calculations for LaFeAsO using the experimental
175 K crystal structure of de la Cruz and co-workers, \cite{cruz}
including the experimental
internal coordinates. We did not relax the internal
coordinates because they are known to be incorrectly given by
non-magnetic GGA calculations, while magnetic calculations incorporating
the spin density wave give overly strong magnetism, presumably due
to spin fluctuations neglected in DFT calculations.
\cite{mazin-mag,singh-ferev}
In any case, this should be kept in mind when comparing calculations,
since the band structure, especially near the zone center, is sensitive
to the As position in the unit cell. In particular, there are additional
hole sheets of Fermi surface that depend on whether one uses the experimental
or calculated As position. Here we focus on the large scale features
of the electronic structure that are not dependent on this choice.
These are the energy dependent character of the bands and the shape of the
main Fermi surface, which governs superconductivity.
\cite{mazin-spm}

The calculated non-spin-polarized DOS, band structure and Fermi surfaces
are shown in Figs. \ref{LaFeAsO-dos}, \ref{LaFeAsO-bands} and
\ref{LaFeAsO-fs}, respectively.
As may be seen, the bands within $\sim$ 2 eV of $E_F$ are of
mainly Fe $d$ character
with modest hybridization of As $p$ states
in both the PBE and TB-mBJ calculations.
The As and O bands are at higher binding energy, yielding nominally Fe$^{2+}$
metallic $d^6$ sheets. Also in both calculations there is a pseudogap in
the DOS starting near $E_F$ corresponding to the semimetallic character.
This is seen in the small disconnected Fermi surfaces.
One difference is that the TB-mBJ calculation has a $\sim$10\% smaller
$d$ band width, which would marginally improve agreement with experiment.
The detailed dispersions of the bands near $E_F$ are different between
the two calculations. The electron pockets are larger in the TB-mBJ
calculation. Also, there is one fewer small hole Fermi surface section.
However, while these features
may be important to understanding the materials,
they are at the level of differences that
arise from different treatments of the As position, and we do
emphasize them here.
The overall agreement of the PBE and TB-mBJ results in this material
is gratifying.

\section{summary and conclusions}

We performed calculations for a diverse set of materials using the
recently developed semilocal density functional of Tran and Blaha.
The present results taken in combination with the tests that they 
reported \cite{mbj} and calculations for halides \cite{singh-halides}
strongly support the conclusion that this functional greatly improves
the band gaps and electronic structure
of simple semiconductors and insulators.
It also greatly improves upon standard density functionals in that
it stabilizes an antiferromagnetic insulating ground state in CaCuO$_2$.
However, the quality of the description of $d$ bands still needs clarification.
In ZnO, the Zn $3d$ states may be too shallow, and in CaCuO$_2$
the distribution the $d$ DOS in the valence bands differs from that
obtained with the standard LDA+U method.
The large scale features of the TB-mBJ electronic structure of LaFeAsO are
similar to those with the PBE functional, although there are changes
in details that affect the hole Fermi surface sections.
Turning to $f$ band materials,
the properties of La$_2$O$_3$ are improved relative to PBE calculations,
although the $4f$ bands are hardly affected at all by the change in
functional. On the other hand, the properties of ferromagnetic Gd are
seriously degraded with respect to the PBE functional.

In summary, the
results indicate that the TB-mBJ functional is a very useful development
for treating the electronic structure of simple semiconductors and
insulators,
including hydrides, while care is needed in applying it to
$d$ and $f$ band materials.

\acknowledgements

Work at ORNL was supported by the Department of Energy, 
ORNL LDRD Program (insulators) and the Office of Basic Energy Sciences,
Materials Sciences and Engineering Division (metals).

\bibliography{tb}

\end{document}